\documentclass{ws-procs975x65}
\def\IR{{\hbox{{\rm I}\kern-.2em\hbox{\rm R}}}}
\def\IC{{\hbox{{\rm I}\kern-.2em\hbox{\rm C}}}}
\def\IR{{\hbox{{\rm I}\kern-.2em\hbox{\rm R}}}}
\def\II{{\hbox{{\rm I}\kern-.2em\hbox{\rm I}}}}
\def\beq{\begin{equation}}
\def\eeq{\end{equation}}
\def\be{\begin{equation}}
\def\ee{\end{equation}}\def\um{\frac{1}{2}}
\def\bea{\begin{eqnarray}}
\def\eea{\end{eqnarray}}
\def\b{\beta}
\def\a{\alpha}

\def\ga{\gamma}
\def\g{\gamma}

\begin{document}

\title{Quantum Holonomies in (2+1)-Dimensional Gravity}

\author{J.~E.~Nelson\footnote{speaker at the conference}}
       
\address{Dipartimento di Fisica Teorica, Universit\`a degli Studi di Torino\\
and Istituto Nazionale di Fisica Nucleare, Sezione di Torino\\
via Pietro Giuria 1, 10125 Torino, Italy\\E-mail: nelson@to.infn.it}

\author{R.~F.~Picken}

\address{Departamento de Matem\'{a}tica and 
CEMAT - Centro de Matem\'{a}tica e Aplica\c{c}\~{o}es \\
Instituto Superior T\'{e}cnico\\Avenida Rovisco Pais, 1049-001 
Lisboa, Portugal\\E-mail: rpicken@math.ist.utl.pt} 


\maketitle

\abstracts{We describe an approach to the quantization of (2+1)--dimensional
gravity with topology $\IR \times T^2$ and negative cosmological constant,
which uses two quantum holonomy matrices satisfying a $q$--commutation
relation. Solutions of diagonal and upper--triangular form are constructed,
which in the latter case exhibit additional, non--trivial {\it internal}
relations for each holonomy matrix. This leads to the notion of quantum matrix
pairs. These are pairs of matrices with non-commuting entries, which have the
same pattern of internal relations, q-commute with each other under matrix
multiplication, and are such that products of powers of the matrices obey the
same pattern of internal relations as the original pair. This has implications
for the classical moduli space, described by ordered pairs of commuting
${\rm SL}(2, \IR)$ matrices modulo simultaneous conjugation by 
${\rm SL}(2, \IR)$ matrices.} 

\section{Introduction}
It is known that the phase space of $(2+1)$-dimensional gravity
with topology
$\IR \times T^2$ and negative cosmological constant $\Lambda$
is described by the space of gauge equivalence classes of flat
$SO(2,2)$ (de Sitter) connections on the torus,
or equivalently by the space of conjugacy classes of homomorphisms from
$\pi_1(T^2)$ to $SO(2,2)$ \cite{wit}.
Since the fundamental group of $T^2$ is generated by two classes
$\gamma_1$ and $\gamma_2$, subject to the relation
$\gamma_1\cdot\gamma_2\cdot\gamma_1^{-1}\cdot\gamma_2^{-1}={\II}$, the
phase space may be identified with pairs of commuting
$SO(2,2)$ elements $(S_1,S_2)$,
identified up to simultaneous conjugation by the same group element
$(S_1,S_2)\sim (g^{-1}S_1g, g^{-1}S_2g)$ for any $g\in SO(2,2)$. These
matrix pairs are the holonomies of the flat connection along the two
generators of the fundamental group.  Using the isomorphism
$\hbox{SO}(2,2)\cong \hbox{SL}(2,\IR)\otimes \hbox{SL}(2,\IR)$ in Ref. \refcite{NRZ} the
Poisson algebra of elements $({U_1}^{\pm},{U_2}^{\pm})$ of
$\hbox{SL}(2,\IR)$ was calculated. It follows from the Poisson brackets
of the connection at points where the curves
intersect. The intersection number  between $\ga_1$ and $\ga_2$ is taken
to be $+1$, and $\gamma_1\cdot\gamma_2$ has intersection number $-1$ with
$\gamma_1$ and $+1$ with $\gamma_2$. The phase space was then
described in terms of the six gauge-invariant (normalized) traces
$~ T_i^{\pm}, i=1,2,3$, where $T_i^{\pm}= {\frac 1 2}{\rm
tr} U_i^{\pm}$, $i=1,2$, $T_3^{\pm}={\frac 1 2}{\rm tr}(U_1^{\pm}U_2^{\pm})$
which satisfy the non--linear cyclical Poisson bracket algebra
\beq
\{T_i^{\pm},T_j^{\pm}\}=\mp{\frac {\sqrt {-\Lambda}} 4}({\epsilon_{ij}}^k 
T_k^{\pm} - T_i^{\pm}T_j^{\pm}), \quad \epsilon_{123}=1
\label{pbr}
\eeq 
where the superscript
$\pm$ refers to the two copies (real and independent) of
$\hbox{SL}(2,\IR)$. The six holonomies $T_i^{\pm}$ of (\ref{pbr}) provide an overcomplete description
of the spacetime geometry of $\IR\!\times\!T^2$.  To see this, consider the 
cubic polynomials
\begin{equation}
F^{\pm}=1-(T_1^{\pm})^2-(T_2^{\pm})^2-(T_3^{\pm})^2 +
 2 T_1^{\pm}T_2^{\pm}T_3^{\pm} .
\label{b9}
\end{equation}
which have vanishing Poisson brackets with all of the
traces $T_i^{\pm}$, are cyclically symmetric in the $T_i^{\pm}$, and
vanish classically by the $\hbox{SL}(2, \IR)$ Mandelstam identities;
setting $F^\pm = 0$ removes the redundancy.

The Poisson algebra (\ref{pbr}) and its generalization \cite{NR3} to more
complicated spatial topologies can be quantized for any value of the
cosmological constant \footnote[1]{There is an analogous discussion for $\Lambda$
positive or zero \cite{NR2}. For example, for $\Lambda$ positive,
the spinor group of the de Sitter
group $\hbox{SO}(3,1)$ is $\hbox{SL}(2,C)$, and the two $\pm$ copies
refer to complex conjugates.}. For a generic topology, one obtains an abstract
quantum algebra \cite{NR1,NR0}.  For genus $1$ with $\Lambda < 0$, the
quantum theory has been worked out explicitly \cite{cn1}.

The holonomies of (\ref{pbr}) can be represented classically as
\beq
T_1^\pm = \cosh{\frac {r_1^\pm} 2} , \quad T_2^\pm  = \cosh{\frac {r_2^\pm} 2} , \quad
T_3^\pm = \cosh{\frac{(r_1^\pm+r_2^\pm)} 2} , \label{cc6}
\eeq
where
$r_{1,2}^{\pm}$ are also real, global, time-independent (but undetermined) 
parameters which, from (\ref{pbr}) satisfy the Poisson brackets
\be\{r_1^\pm,r_2^\pm\}=\mp \sqrt {-\Lambda}, \qquad \{r_{1,2}^+,r_{1,2}^-\}= 0.
\label{pb}
\ee
In this case the cubic polynomials (\ref{b9}) are identically zero, 
and quantization \footnote[2]{Direct quantization of the algebra (\ref{pbr})
gives an algebra related to the Lie algebra of the quantum group
$\hbox{SU}(2)_q$ \cite{NRZ,NR5}, where $q=\exp{(4i\theta)}, \tan\theta=
- {\frac {\hbar \sqrt {-\Lambda}} 8}$. A (scaled) representation of the operators (\ref{cc6}) leads to
the commutators $[\hat r_1^{\pm}, \hat r_2^{\pm}] = \pm 8i\theta$, which differ
from (\ref{dc1}) by terms of order $\hbar^3$.}  is achieved by replacing
Eqs.(\ref{pb}) with the commutators
\begin{equation}
[\hat r_1^\pm, \hat r_2^\pm] = \mp  i\hbar \sqrt {-\Lambda}, \quad [\hat r_{1,2}^+, 
\hat r_{1,2}^-] = 0 .
\label{dc1}
\end{equation}

Previous quantizations \cite{cn1,cn2} have
concentrated entirely on the traces $T_i^{\pm}$ and their
representation Eq.(\ref{cc6}). Here we observe that we may
regard the quantized traces $\hat T_i^\pm$ as traces of 
operator-valued holonomy matrices $\hat T_i^{\pm}= \um {\rm tr}
\hat U_i^{\pm}$, $i=1,2$, $\hat T_3^{\pm}= \um {\rm tr}(\hat
U_1^{\pm} \hat U_2^{\pm})$, where (for the (+)  matrices, dropping
the superscript) the matrices $\hat U_i$ have, for example, the diagonal form 
\beq \hat U_i
= \left(\begin{array}{clcr}e^{\frac {{\hat r}_i} 2}&~~0\\0& e^{-{\frac {{\hat
r}_i} 2}} \end{array}\right)\quad 
= \exp{({\frac {\hat r_i \sigma_3} 2})} \label{diag} 
\eeq 
where $\sigma_3$ is one of the Pauli matrices. Now, from Eq.(\ref{dc1}) and the 
identity
$$ e^{\hat X}
e^{\hat Y}= e^{\hat Y} e^{\hat X} e^{[ \hat X, \hat Y ]},
$$
valid when
$[ \hat X, \hat Y ]$ is a $c$--number, one finds that the 
matrices (\ref{diag}) satisfy, {\it by both matrix and operator
multiplication}, the $q$--commutation relation\footnote[3]{the (-) matrices
satisfy a similar relation but with $q$ replaced by $q^{-1}$}:
\beq
\hat U_1 \hat U_2 = q \hat U_2 \hat U_1, \quad {\rm with} 
\label{fund}
\eeq
\beq
q=\exp (-{\frac {i \hbar \sqrt{-\Lambda}}  4})
\label{q}
\eeq
i.e. a deformation of the classical equation stating that the holonomies
commute.

Equations of the form (\ref{fund}) appear abundantly in the quantum group and
quantum geometry literature, as Weyl relations or $q$--commutators, or as the
defining relation for the quantum plane \cite{man}, but normally the symbols
$\hat U_1$ and $\hat U_2$ stand for scalar operators, as opposed to $2\times 2$
matrices with operator entries. 

Our approach is based on the fundamental equation
(\ref{fund}). Instead of representing the algebra of traces, Eq.
 (\ref{pbr}), we find representations of matrices $\hat U_1$ and $\hat
U_2$ satisfying Eq. (\ref{fund}) that generalize the choices
(\ref{diag}), for a general $q$--parameter. This constitutes a new
approach to quantization that is consistent with previous
approaches for this model \cite{NR1,cn1,cn2}, namely a deformation
of classical holonomies that consequently satisfy a
$q$--commutation relation. The gauge-invariance of
the traces is replaced by the gauge-covariance of Eq. (\ref{fund}) under
the replacements $\hat U_i\rightarrow g^{-1}\hat U_ig$, $i=1,2$ for
$g\in \hbox{SL}(2,\IR )$ an ordinary, i.e. not operator-valued,
matrix. We argue that working directly with the matrices $U_i$,
rather than with the indirect information contained in their
traces, gives a clearer insight into the structure of the phase
space, both classically and after quantization.

A more detailed account of these results, including representations, is given
in Ref. \refcite{NP1}. The matrices $\hat U_1$ and $\hat U_2$ determine a new
quantum--group--like structure, which is studied from the algebraic perspective
in Ref. \refcite{NP2}. The description of the classical phase space in terms of
pairs of matrices $U_i$ is given in Ref. \refcite{NP3}. The generalization to supergroups in the context of (2+1)--supergravity is described in Ref. \refcite{mik:pic}. 

\section{An algebraic solution}
We give just one example of a purely algebraic solution to Eq. (\ref{fund}). For
others see Ref. \refcite{NP2}. Consider the upper--triangular matrices 
\beq
U_i=\left( \begin{array}{cc} \a_i &\b_i\\ 0&\g_i\end{array}
\right), i=1,2
\label{Ui}
\eeq
which generalize Eq. (\ref{diag}). It can be checked that they will satisfy
Eq. (\ref{fund}) provided their non--commuting elements satisfy the following
mutual relations 
\be
\a_1\a_2=q\a_2\a_1,\quad \g_1\g_2=q\g_2\g_1, \quad {\rm and}
\label{ab1}
\ee 
\be
 \a_1\b_2 = q\b_2\g_1, \quad \b_1\g_2 = q \a_2\b_1
\label{ab11}
\ee
and the following {\it internal} relations for each ($i=1,2$) matrix
\be
\a_i\g_i=\g_i\a_i=1,\quad \a_i\b_i=\b_i\g_i \label{int}
\ee

Note that the mutual relations (\ref{ab1}) are standard $q$--commutation 
relations \cite{man}, which clearly become commutative in the classical limit 
$q \to 1$, whereas relations (\ref{ab11}) have a different structure, involving 
three elements not two. The relations (\ref{ab1}) also imply that, for example
\be
\a_1{\a_2}^{-1}=q^{-1}{\a_2}^{-1}\a_1,
\quad {\a_1}^{-1}{\a_2}^{-1}=q{\a_2}^{-1}{\a_1}^{-1},
\label{ab12}
\ee
and similarly for $\g_1,\g_2$. The internal relations Eqs. (\ref{int}) are a 
new feature, since, in the Poisson algebra of (2+1)--dimensional gravity
\cite{NRZ}, only matrix elements from different holonomies have
non--zero brackets and would therefore not commute on quantization.
Elements of a single holonomy commute.  Note that the $q$
parameter does not appear in the internal relations (\ref{int})
which therefore persist in the classical limit $q\rightarrow 1$,
when the matrices (\ref{Ui}) commute.

The internal relations (\ref{int}) are also not standard $q$--commutation
relations, but are preserved under matrix multiplication,
and in this sense they are analogous to the internal relations for quantum
groups \cite{man}. For example, the product $U_1U_2$ is given by
\be
U_1U_2 = \left(\begin{array}{clcr}{\a_1\a_2}&{~~\a_1\b_2 + \b_1\g_2}\\
0&{~~\g_1\g_2}
\end{array}\right)
\label{matrvv}
\ee
whose internal relations are analogous to Eq. (\ref{int}), by using 
(\ref{ab1})--(\ref{ab11}) and (\ref{int}). This feature is discussed in 
greater detail in Ref. \refcite{NP1}.

\section{The classical moduli space} 
The new, non--trivial {\it internal} relations, Eq. (\ref{int}) have important
implications for the classical phase space, which consists of pairs of 
commuting $\hbox{SL}(2,\IR)$ matrices, identified up to simultaneous
conjugation by elements of $\hbox{SL}(2,\IR)$, since the classical counterpart
to Eq. (\ref{fund}) is the statement that the two matrices $U_1$ and $U_2$
commute. This space was studied in detail in Ref. \refcite{eza}. The algebraic analysis
of the classical case, in terms of the eigenvalues and eigenspaces of the two
matrices, does not carry over in any straightforward way to quantum matrices.
For instance, an upper--triangular matrix with two distinct diagonal entries
can be diagonalized as an ordinary matrix, but this is not in general true if
the matrix has non--trivial {\it internal} relations between non--commuting
entries. In Ref. \refcite{NP3} we give another parametrization of the classical
phase space, which is more appropriate to the present context. It consists of
sectors where both matrices are diagonalizable, but also sectors where both are
non-diagonalizable but can be simultaneously conjugated into upper triangular
form, as well as other sectors. A spectral analysis of commuting
$\hbox{SL}(2,\IR)$ matrices allows a classification of the equivalence classes,
and a unique canonical form is given for each of these. In this way the moduli
space becomes explicitly parametrized, and has a simple structure, resembling
that of a cell complex, allowing it to be depicted. Full details are given in
Ref. \refcite{NP3}.

\section*{Acknowledgments} This work was supported by the Istituto Nazionale
di Fisica Nucleare (INFN) of Italy, Iniziativa Specifica FI41, the Italian
Ministero dell'Universit\`a e della Ricerca Scientifica e Tecnologica (MIUR),
and by the programme {\em Programa Operacional 
``Ci\^{e}ncia, Tecnologia, Inova\c{c}\~{a}o''} (POCTI) of the 
{\em Funda\c{c}\~{a}o para a Ci\^{e}ncia e a Tecnologia} (FCT), 
cofinanced by the European Community fund FEDER.

\end{document}